\documentclass[prl,twocolumn,showpacs,preprintnumbers]{revtex4}
\usepackage{graphicx}
\newcommand{\Vec}[1]{\mbox{\boldmath$#1$}}
\begin{document}

\preprint{to be published in Phys. Rev. Lett.}

\title{Off-site-repulsion induced Triplet Superconductivity\\ 
--- a possibility of chiral p$_{x+y}$ pairing in Sr$_2$RuO$_4$ 
}

\author{
Ryotaro Arita$^1$, Seiichiro Onari$^1$, 
Kazuhiko Kuroki$^2$, and Hideo Aoki$^1$
}

\affiliation{$^1$Department of Physics, University of Tokyo, Hongo,
Tokyo 113-0033, Japan}
\affiliation{$^2$Department of Applied Physics and Chemistry,
University of Electro-Communications, Chofu, Tokyo 182-8585, Japan}

\date{\today}

\begin{abstract}
In order to probe the effect of charge fluctuations on triplet pairing, 
we study the pairing symmetry in the one-band Hubbard model 
having the off-site Coulomb repulsion ($V$) on top of the on-site repulsion 
as a model for the $\gamma$ band of Sr$_2$RuO$_4$, a strong candidate 
for triplet pairing superconductor.
The result, obtained with the dynamical cluster 
approximation combined with the quantum Monte-Carlo method, 
and confirmed from the fluctuation exchange approximation, 
shows that while $d_{x^2-y^2}$-pairing dominates over $p$ 
in the absence of $V$, 
introduction of $V$ makes $p_{x+y}$ and $d_{xy}$ 
dominant. The gap function for the chiral $p_{x+y}$+i$p_{x-y}$ 
has nodes that are consistent with 
the recent measurement of specific heat in rotated magnetic fields 
in the ruthenate.  
This suggests that the off-site repulsion may play 
an essential role in triplet superconductivity in this material. 
\end{abstract}
\maketitle

{\it Introduction} --- 
There has been an increasing fascination with 
spin-triplet pairing in the condensed matter physics, with a history 
dating back to the discovery of superfluid $^3$He.
While triplet superconductivity has been found in some heavy 
fermion compounds and organic metals, a ruthenate, Sr$_2$RuO$_4$, 
is prototypical in that its structure is similar to a cuprate La$_2$CuO$_4$,
but the replacement of Cu with Ru makes the relevant d orbitals different.  
So the discovery of superconductivity\cite{Maeno94} 
in the ruthenate has stimulated an enormous amount of 
studies\cite{Maeno03}. 
Experimentally, it has been established that Sr$_2$RuO$_4$ is a 
chiral (time-reversal broken) triplet superconductor 
with the spins lying in the RuO$_2$ plane, where
we take the spin quantization axis ($z$) along the crystalline 
$c$-axis, as indicated from Knight shift\cite{Ishida98} and 
$\mu$SR\cite{Luke98} experiments.

On the other hand, the symmetry of the gap function,
which is of prime importance in identifying the pairing mechanism,
has yet to be established. Although the existence of line-nodes in the 
gap function is suggested from power-law behaviors in 
specific heat and NMR 1/$T_1 T$\cite{Maeno03}, 
the position of nodes remains controversial.
While the magnetothermal conductivity shows a weak anisotropy
in the $ab$-plane for $T>0.35K$\cite{Tanatar01,Izawa01}, four-fold 
oscillations [indicative of nodes in the gap function 
around $\Vec{k}=$($\pm \pi$,0), (0,$\pm \pi$)] have been detected 
in a recent measurement of specific 
heat in rotating magnetic fields by Deguchi {\it et al.}\cite{Deguchi03}. 

One complication is that, 
unlike the high-Tc cuprates which have a square lattice 
of d$_{x^2-y^2}$ orbitals, the ruthenate 
has three, cylindrical Fermi surfaces (labelled as $\alpha$, $\beta$, 
and $\gamma$), where $\alpha$ and $\beta$ derive 
from one-dimensional arrays of 
Ru $d_{xz}$ and $d_{yz}$ orbits, while $\gamma$ 
derives from a square lattice of $d_{xy}$ orbits. 
This has been established experimentally from 
a de Haas measurement\cite{Bergemann02} 
and an angle-resolved photoemission 
spectroscopy,\cite{Damaselli00} and also 
confirmed from a first-principles band calculation\cite{Oguchi95}.   
Deguchi {\it et al.} have concluded that their result suggests that 
the active band for superconductivity is the $\gamma$ band.\cite{Deguchi03}

Although various pairing mechanisms have been 
proposed for Sr$_2$RuO$_4$, they are not straightforward, since 
it is much more difficult to explain triplet pairing 
than singlet pairing solely from spin fluctuations 
because of a smaller pairing 
interaction in the triplet channel. Several 
authors\cite{Kuwabara00,Sato00,Takimoto00,Kuroki02} have 
focused on the effect of nesting in the quasi-one-dimensional 
$\alpha$-$\beta$ Fermi surfaces to show that 
anisotropic antiferromagnetic spin fluctuations, observed 
in NMR experiment\cite{Mukuda98}, or orbital fluctuations 
can favor the triplet pairing.
On the other hand, assuming that the $\gamma$ band is the active band
for superconductivity, a one-loop renormalization group analysis 
for the one-band Hubbard model was performed\cite{Honerkamp01}, 
where $p$-pairing is concluded to be dominant unless the Fermi level is 
far from the van Hove singularity.
A third-order perturbation calculation for
the single-band Hubbard model has also been 
performed{\it et al.}\cite{Nomura00,Yanase03}, where
triplet pairing is shown to dominate over singlet pairing 
for intermediate band filling. 
This was recently extended to the three-band Hubbard 
model,\cite{Nomura02,Yanase03} for which the perturbation 
calculation shows that triplet pairing remains dominant, 
residing on the $\gamma$ band. 
Their result that the gap function has nodes along [100] for the $\gamma$ band, 
while they lie along [110] for $\alpha$-$\beta$, 
is consistent with the anisotropy in the magnetothermal 
conductivity\cite{Tanatar01,Izawa01}. However, we believe that
the validity of the results obtained with perturbation expansions, 
truncated at the third 
or fourth order, has to be checked by non-perturbative methods.

Indeed, according to the fluctuation exchange (FLEX) 
study\cite{Arita99} or a phenomenology\cite{Monthoux99} 
for the spin fluctuation, triplet pairing is 
rather weak in general.  More specifically, Kuroki {\it et al.} 
have shown recently that the singlet pairing dominates over the triplet 
for the one-band Hubbard model for 
the $\gamma$ band, where they 
calculated the pairing interaction vertex with the quantum Monte Carlo 
(QMC) method to show that $d$-pairing interaction is stronger 
than those in triplet channels.\cite{Kuroki03}

Given this background, the motivation of the present study is two-fold. 
First, we want to clarify whether 
the triplet superconductivity can be dominant within the one-band 
Hubbard model in a {\it non-perturbative} method, for which we have employed 
the dynamical cluster approximation (DCA) 
combined with QMC formulated by 
Jarrell {\it et al.}\cite{Jarrell}. 
We shall show that $d_{x^2-y^2}$-wave pairing dominates over 
$p$-wave pairings suggested by third-order perturbation studies.
Here DCA+QMC is employed for the first time\cite{comment1}
to investigate superconductivity in the extended Hubbard Hamiltonian 
that models the $\gamma$ band in the ruthenate. 

Second, and more importantly, 
we propose that the {\it off-site Coulomb repulsion} can 
favor triplet superconductivity in general and in 
the $\gamma$ band in Sr$_2$RuO$_4$ in particular\cite{comment5}. 
The underlying physics is that, while we usually evoke 
spin fluctuations (caused by the on-site repulsion) 
in considering an electron mechanism of superconductivity, 
charge fluctuations, which tend to be enhanced by 
the off-site repulsion, should favor the pairing in the triplet channel, 
as seen in the expression for the fluctuation-mediated interactions 
\cite{Takimoto00,Kuroki01}.
We have in fact shown with a FLEX calculation that, 
even for the simple square lattice, 
transitions between different pairing symmetries can arise due 
to the coexistence of spin and charge fluctuations\cite{Onari03}.
We shall show that the off-site repulsion makes $p_{x+y}$ and $d_{xy}$ 
dominant while $p_{x}$ and $d_{x^2-y^2}$ are suppressed. The result 
is also confirmed by a FLEX calculation. 
The chiral $p_{x+y}$+i$p_{x-y}$ gap function then 
has nodes around ($\pm \pi$,0) and (0,$\pm \pi$), which is consistent 
with the specific heat in rotating magnetic fields\cite{Deguchi03}. 

{\it Formulation} --- The extended Hubbard model is given as 
\begin{eqnarray*}
{\cal H}&=&-t\sum_{\langle i,j \rangle,\sigma}^{\rm nn}
(c_{i\sigma}^{\dagger}c_{j\sigma} + {\rm h.c.})
-t'\sum_{\langle i,j \rangle,\sigma}^{\rm nnn}
(c_{i\sigma}^{\dagger}c_{j\sigma} + {\rm h.c.})\\
&&+U\sum_{i}n_{i\uparrow}n_{i\downarrow}\nonumber
+V\sum_{\langle i,j\rangle,\sigma\sigma'}^{\rm nn}
n_{i\sigma}n_{j\sigma'}
\end{eqnarray*}
in the standard notation, where nn (nnn) denotes 
nearest-neighbor (next-nn) sites. 
In DCA, the original reciprocal space with $N$ points are divided 
into $N_c$ cells, for which a coarse-graining is 
done. The Hubbard model is then mapped to a self-consistently embedded
cluster of Anderson impurities (rather than a single impurity considered
in the dynamical mean-field approximation\cite{DMFA}), so that
DCA incorporates nonlocal spatial fluctuations.  
We have solved the cluster problem generated by the DCA using
the QMC with the algorithm proposed by Hirsch and Fye\cite{Hirsch}. 
We choose a cluster size $N_c=4\times 4$ throughout the study\cite{comment2}.

The effect of the off-site repulsion 
$V$ can be incorporated into the DCA calculation as follows. 
The off-site term is expressed in $k$-space as
\begin{eqnarray*}
\sum_{\langle i,j\rangle,\sigma\sigma'}^{\rm nn}
V_{ij}n_{i\sigma}n_{j\sigma'}=\frac{1}{N^2}
\sum_{\Vec{k} \Vec{k}' \Vec{q}} V_{\Vec{q}}
c_{\Vec{k}\sigma}^{\dagger}c_{\Vec{k}
+\Vec{q}\sigma}c_{\Vec{k}'\sigma}^{\dagger}c_{\Vec{k}'-\Vec{q}\sigma}  
\end{eqnarray*}
with $V_{\Vec{q}}=(V/N)[\cos (q_x)+\cos (q_y)]$, which has to be 
coarse-grained.  
One might think that the coarse-graining may introduce 
interactions extending beyond nearest neighbors for the impurity
cluster model. However, we can note a relation, 
$\tilde{V}_{\Vec{Q}} \equiv (N_c/N)\sum_{\tilde{\Vec{q}}}
V_{\Vec{Q}+\tilde{\Vec{q}}} 
= (\tilde{V}/N_c)(\cos Q_x +\cos Q_y)$,
where the summation for $\tilde{\Vec{q}}$ runs over 
the momenta in the coarse-graining cell centered at $\Vec{Q}$, and 
$\tilde{V} \equiv \sin(\pi/N_c)/(\pi/N_c)V$.  
Thus the form of the interaction does not change, 
so we have only to consider the off-site repulsion $\tilde{V}$ 
in the QMC for the impurity cluster model.

\begin{figure}[t]
\begin{center}
\includegraphics[width=5cm,clip]{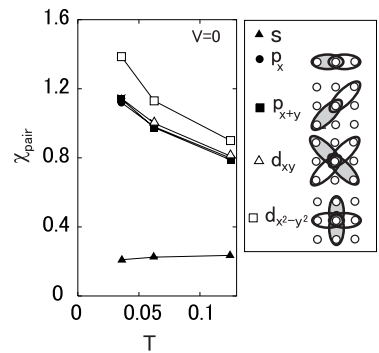}
\end{center}
\caption{(a) DCA+QMC result for the pairing susceptibility versus temperature 
for various symmetries of pairing (depicted in real space in inset) 
in the on-site Hubbard model ($V=0$) 
with the on-site interaction $U=1.5$, the band filling 
$n=1.33$, and $t'/t=0.4$. Error bars are smaller than the size of
each symbol.  
}
\label{fig1}
\end{figure}

{\it Results} ---  
In the DCA+QMC calculation, 
we take the half of the band width as the unit of energy as customary done. 
We take the on-site Coulomb interaction $U=1.5$, the band filling $n=1.33$, 
and $t'/t=0.4$.  These values are chosen to roughly represent the 
$\gamma$ band of the ruthenate.
The pairing symmetries considered are 
$s \sim 1$, $p_x \sim \sqrt{2}\sin (k_x)$, 
$p_{x\pm y} \sim \sqrt{2}\sin (k_x \pm k_y)$,\cite{commentpxy} 
$d_{x^2-y^2} \sim \cos (k_x)-\cos (k_y)$, and 
$d_{xy} \sim \cos (k_x+k_y)-\cos (k_x-k_y)$, as depicted in real space 
in the inset of Fig.\ref{fig1}.  
We first show the result before $V$ is turned on. Figure \ref{fig1} 
shows the pairing susceptibility for the on-site Hubbard model 
as a function of temperature, where 
$L=64$ Trotter-Suzuki decomposition number is taken.  
We can see that $d_{x^2-y^2}$ is the dominant pairing, 
while $p_{x+y}$ is weaker 
in the region $T \geq 0.031$ studied here. 
The result is consistent with Ref.\cite{Kuroki03},
where $d_{x^2-y^2}$ is shown to dominate over 
triplet pairings by a QMC for a finite ($N=14\times 14$) 
Hubbard model with $U=0.5$. 
These results, obtained with non-perturbative methods, 
suggest that the on-site Hubbard model is 
insufficient to describe the superconductivity in Sr$_2$RuO$_4$. 

\begin{figure}[t]
\begin{center}
\includegraphics[width=5cm,clip]{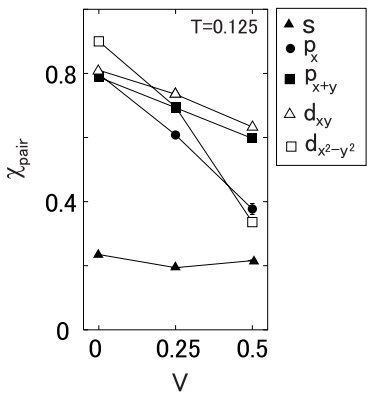}
\end{center}
\caption{The pairing susceptibility in the extended Hubbard model
as a function of the off-site repulsion $V$ for 
$U=1.5$, $n=1.33$, and $t'/t=0.4$ at $T=0.125$.
Error bars are smaller than the size of each symbol.}
\label{fig2}
\end{figure}

Now we move on to the extended Hubbard model in Fig.\ref{fig2}, 
where the DCA+QMC result for the 
pairing susceptibility is plotted as a function of 
the off-site interaction $V$ for $T=0.125$ 
with the Trotter-Suzuki decomposition number $L=24$.
We can clearly see a qualitative tendency that 
$p_{x+y}$ and $d_{xy}$ become dominant with the introduction of 
$V$, while $p_{x}$ and $d_{x^2-y^2}$ 
are suppressed. Physically, this should be because  
the nearest-neighbor $V$ suppresses pairs formed across 
nearest-neighbor sites, while 
the pairs such as $p_{x+y}$ and $d_{xy}$ 
that are formed across more distance sites (see the inset of Fig.\ref{fig1}) 
are less affected.
We cannot take a lower $T$ or larger $L$ at present, 
since in the QMC algorithm for 
$V\sum n_{i\sigma}n_{i\sigma'}$ eight Hubbard-Stratonovich 
auxiliary fields (for $\sigma,\sigma'=\uparrow,\downarrow$ for 
$x,y$ directions) are needed 
on top of the one for the $U$-term, so the study of the subtle 
competition between 
$p_{x+y}$ and $d_{xy}$ at lower temperatures is a future study\cite{comment3}. 

\begin{figure}[t]
\begin{center}
\includegraphics[width=6cm,clip]{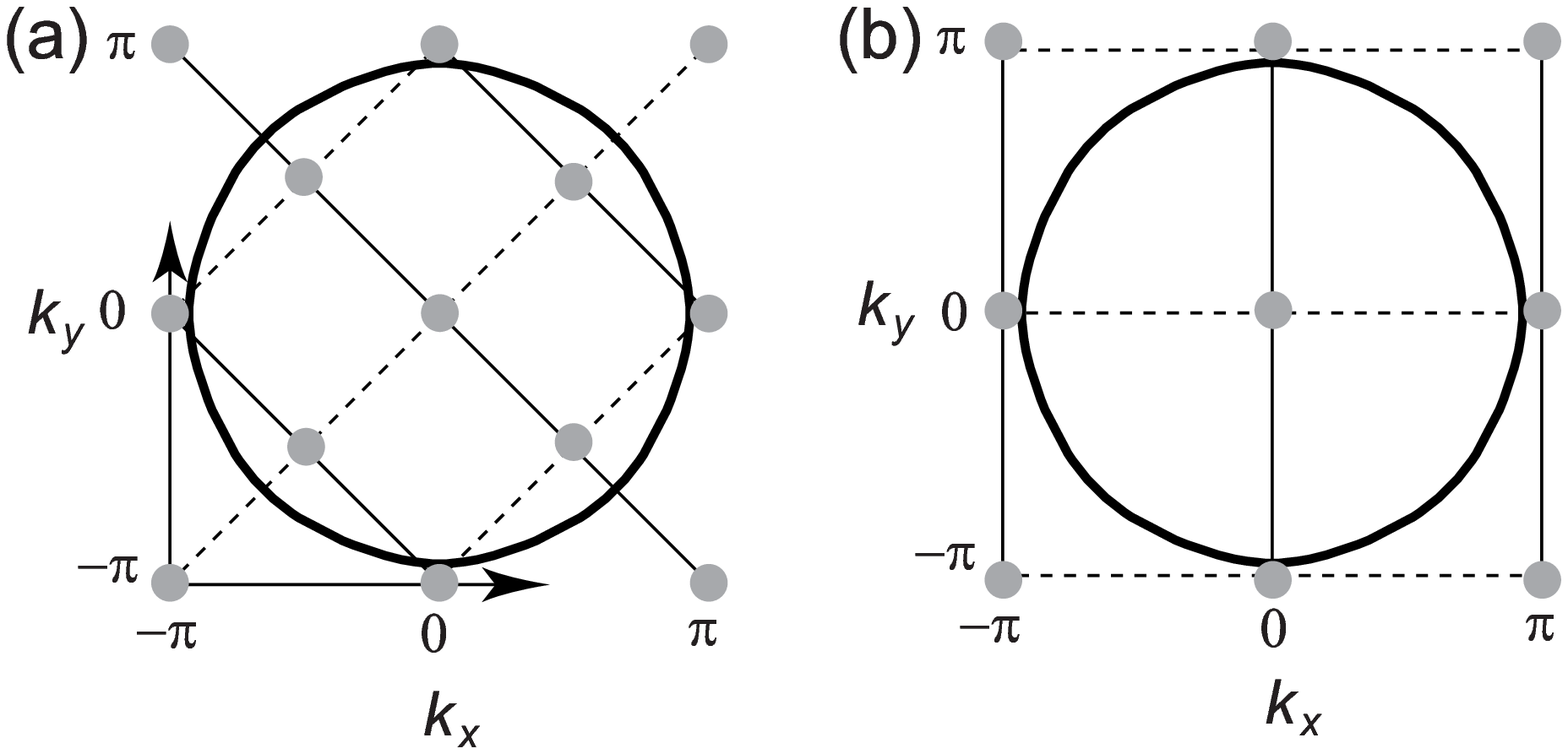}
\end{center}
\caption{(a) The position of nodes 
of $p_{x+y} \sim \sin(k_x+k_y)$ (solid lines),
$p_{x-y} \sim \sin(k_x-k_y)$ (dotted), and 
$p_{x+y}+ip_{x-y}$ (gray dots). 
The bold solid curve represents the $\gamma$ Fermi surface 
of Sr$_2$RuO$_4$. (b) A plot similar to (a) 
for $p_{x}+ip_{y} \sim \sin(k_x)+i \sin(k_y)$.}
\label{fig3}
\end{figure}

Still, it is interesting to discuss what should result from the 
$p$-wave pairing.  Below $T_c$ the $p_{x+y}$-wave pairing is 
expected to take a chiral form, 
\[
p_{x+y}+ip_{x-y} \sim \sin (k_x + k_y) + i\sin (k_x - k_y),
\]
since the mixing should increase the gap energy ($|\Delta|$).
We can then note, as depicted in Fig.\ref{fig3}, 
that the gap function $\Delta$ for the $p_{x+y}+ip_{x-y}$ 
has nodes (or minima of $|\Delta|$ on the Fermi surface) 
at around ($\pm \pi$,0) and (0,$\pm \pi$) (which happen to be 
similar to those for 
the chiral $p_{x}+ip_{y} \sim \sin(k_x)+i \sin(k_y)$; Fig.\ref{fig3}(b)).  
So the chiral $p_{x+y}+ip_{x-y}$ is consistent with the specific heat 
measurement\cite{Deguchi03}, and should be a 
candidate for the triplet superconductivity in Sr$_2$RuO$_4$. 

We finally examine whether the FLEX, a renormalized perturbation, 
would reproduce the above result.  
The FLEX was first formulated by Bickers {\it et al.}\cite{Bickers89}
for the Hubbard model, and further applied to the extended Hubbard model 
by Esirgen {\it et al.}\cite{Esirgen97}.
We first obtain the renormalized Green's function, $G$, taking
bubble and ladder diagrams as the self energy.
We then calculate the pairing interaction mediated by spin 
and charge fluctuations, and plug that in Eliashberg's equation, 
\begin{eqnarray}
\lambda\phi(k)&=&
-\frac{T}{N}\sum_{k'}\Gamma^{\rm pp}(k-k')G(k')G(-k')\phi(k')
\label{elia},
\end{eqnarray}
where $\phi$ is the gap function, $k\equiv (\Vec{k},\omega_n)$ 
with $\omega_n$ being Matsubara frequency, 
and $\Gamma^{\rm pp}$ 
the pairing interaction between the pairs with 
$(\Vec{k},-\Vec{k})$ and $(\Vec{k}',-\Vec{k}')$. 

The maximum eigenvalue $\lambda$ becomes unity 
when $T$ becomes the transition temperature of the dominant pairing. 
We take $N=32\times 32$ sites at temperature 
$T=5\times 10^{-3}$ and $-(2N_c-1)\pi T \leq \omega_n \leq (2N_c-1)\pi T$ 
with $N_c=1024$.
When the off-site interaction $V$ is introduced 
all the vertices and susceptibilities become 
$(Z+1)\times (Z+1)$ matrices for the lattice coordination number $Z 
(=4$ for the square lattice).

\begin{figure}[t]
\begin{center}
\includegraphics[width=5cm,clip]{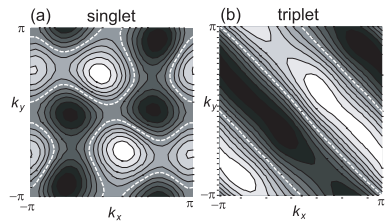}
\end{center}
\caption{Contour plot of the gap function for
the singlet(a) and triplet(b) pairs obtained with 
FLEX+Eliashberg's equation for the extended Hubbard model for 
$U=2.5$, $V=0.65$, $n=1.33$, $t'/t=0.5$, and 
$T=5\times 10^{-3}$. White dashed lines denote the nodes.  
For (b), another one, rotated by 90 degrees, is degenerate.
}
\label{fig4}
\end{figure}

We have performed the FLEX calculation for
$t'/t=0.4, 0.5$, $n=1.33$, $U=1.5 - 2.5$, $V=0 - U/4$.
While the singlet pairing is dominant in most of the 
parameter region, we have found that the maximum eigenvalue of
Eliashberg's equation becomes greater for triplet pairing
than that of the singlet for $0.625 \leq V \leq 0.675$ 
for $U=2.5$\cite{comment4}. 
Figure \ref{fig4} depicts the gap functions for
triplet and singlet cases. 
The symmetry of the gap function, which is 
$p_{x+y}$- and $d_{xy}$-like, respectively, does agree 
with the above DCA+QMC result.\cite{commentd}  

In summary, we have found using DCA+QMC and FLEX methods that, 
while $p$-wave pairing is weaker than $d_{x^2-y^2}$ 
in the on-site Hubbard model, introduction of the 
off-site repulsion $V$ suppresses
$p_{x}$ and $d_{x^2-y^2}$, making 
$p_{x+y}$ and $d_{xy}$ more favorable.  
Although the effect of $\alpha$ and $\beta$ bands is 
beyond the scope of the present study,
the qualitative tendency of the present work that
the off-site repulsion acts to preferentially
suppress nearest neighbor pairs
should hold for multibands as well as for single bands.
The position of the nodes in the 
gap function for the chiral $p = p_{x+y}+ip_{x-y}$ 
is consistent with a specific heat measurement for 
Sr$_2$RuO$_4$\cite{Deguchi03},  
so this is a promising candidate for the 
triplet superconductivity in Sr$_2$RuO$_4$, and
the off-site repulsion may play an essential role there.   

This work was supported by a Grant-in-aid for scientific
research and Special coordination funds from the ministry of
education of Japan. Numerical calculations were performed on
SR8000 in ISSP, University of Tokyo.


\begin{references}
\bibitem{Maeno94}
Y. Maeno, H. Hashimoto, K. Yoshida, S. Nishizaki, T. Fujita, 
J. G. Bednortz, and F. Lichtenberg, Nature {\bf 372}, 532 (1994).
\bibitem{Maeno03}
A. P. Mackenzie and Y. Maeno, Rev. Mod. Phys. {\bf 75}, 657 (2003), 
and refs therein.
\bibitem{Ishida98}K. Ishida, H. Mukuda, Y. Kitaoka, K. Asayama,
Z.Q. Mao, Y. Mori, and Y. Maeno, Nature {\bf 396}, 658 (1998).
\bibitem{Luke98}G.M. Luke, Y. Fudamoto, K. M. Kojima, M. I. Larkin,
J. Merrin, B. Nachumi, Y.J. Uemura, Y. Maeno, Z.Q. Mao, Y. Mori,
H. Nakamura, and M. Sigrist, Nature {\bf 394}, 558 (1998).
\bibitem{Tanatar01}M. A. Tanatar, M. Suzuki, S. Nagai, Z.Q. Mao,
Y. Maeno, and T. Ishiguro, Phys. Rev. Lett. {\bf 86}, 2649 (2001).
\bibitem{Izawa01}K. Izawa, H. Takahashi, H. Yamaguchi, Y. Matsuda, M. Suzki,
T. Sasaki, T. Fukase, Y. Yoshida, R. Settai, and Y. Onuki,
Phys. Rev. Lett. {\bf 86}, 2653 (2001).
\bibitem{Deguchi03}K. Deguchi, Z. Q. Mao, H. Yaguchi, and Y. Maeno,
Phys. Rev. Lett. {\bf 92} 047002 (2004).
\bibitem{Bergemann02} C. Bergemann, A.P. Mackenzie, S.R. Jullian,
D. Forsythe, and E. Ohmichi, Adv. Phys. {\bf 52}, 639 (2002).
\bibitem{Damaselli00}A. Damascelli, K.M. Shen, D.H. Lu, N.P. Armitage,
F. Ronning, D. L. Feng, C. Kim, Z.-X. Shen, T. Kimura, Y. Tokura, 
Z. Q. Mao, and Y. Maeno, 
J. Electron Spectr. Relat. Phenom. {\bf 114}, 641 (2001).
\bibitem{Oguchi95} T. Oguchi, Phys. Rev. B {\bf 51}, 1385 (1995).
\bibitem{Kuwabara00} T. Kuwabara and M. Ogata, Phys. Rev. Lett.
{\bf 85}, 4586 (2000).
\bibitem{Sato00} M. Sato and M. Kohmoto, J. Phys. Soc. Jpn.
{\bf 69}, 3505 (2000).
\bibitem{Takimoto00} T. Takimoto, Phys. Rev. B {\bf 62}, 14641(R) (2000).
\bibitem{Kuroki02} K. Kuroki, M. Ogata, R. Arita, and H. Aoki,
Phys. Rev. B {\bf 63}, 60506(R) (2002). 
\bibitem{Mukuda98} H. Mukuda, K. Ishida, Y. Kitaoka, K. Asayama,
Z. Q. Mao, Y. Mori and Y. Maeno, J. Phys. Soc. Jpn., {\bf 67}
3945 (1998).
\bibitem{Honerkamp01}C. Honerkamp and M. Salmhofer, Phys. Rev. Lett. 
{\bf 87}, 187004 (2001); Phys. Rev. B {\bf 64}, 184516 (2001). 
\bibitem{Nomura00} T. Nomura and K. Yamada, J. Phys. Soc. Jpn.
{\bf 69}, 3678 (2000).
\bibitem{Yanase03} Y. Yanase, T. Jujo, T. Nomura, H. Ikeda, T. Hotta, 
and K. Yamada, Phys. Rep. {\bf 387}, 1 (2003).
\bibitem{Nomura02} T. Nomura and K. Yamada, J. Phys. Soc. Jpn. 
{\bf 71}, 404 (2000).
\bibitem{Arita99} R. Arita, K. Kuroki, and H. Aoki, Phys. Rev. B 
{\bf 60}, 14585 (1999); J. Phys. Soc. Jpn. {\bf 69}, 1181 (2000).
\bibitem{Monthoux99} P. Monthoux and G. Lonzarich, Phys. Rev. B
{\bf 59}, 14598 (1999).
\bibitem{Kuroki03} K. Kuroki, Y. Tanaka, T. Kimura, and R. Arita,
cond-mat/0307553.
\bibitem{Jarrell}M.H. Hettler, A.N. Tahvildar-Zadeh, M. Jarrell, 
T. Pruschke, and H.R. Krishnamurthy, Phys. Rev. B {\bf 58}, (1998) 7475;
M.H. Hettler, M. Mukherjee, M. Jarrell, and H.R. Krishnamurthy, 
Phys. Rev. B {\bf 61}, 12739 (2000); Th. Maier, M. Jarrell, T. Pruschke, 
and J. Keller, Euro. Phys. J. B {\bf 13}, (2000) 613; M. Jarrell, 
Th. Maier, C. Huscroft, and S. Moukouri, Phys. Rev. B {\bf 64}, 195130 (2001). 
\bibitem{comment5}
In general, the magnitude of $V$ is substantial in transition 
metal oxides. For example, $V \simeq U/4$ for cuprates
(see e.g., G. Esirgen, H.B. Schuttler and N.E. Bickers 
[Phys. Rev. Lett. {\bf 82} 1217 (1999)]), and
$V$for Ru 4d is expected to be larger than that for Cu 3d,
because 4d orbitals spread more extensively than 3d orbitals.
\bibitem{Kuroki01}K. Kuroki, R. Arita, and H. Aoki,
Phys. Rev. B {\bf 63}, 94509 (2001). 
\bibitem{Onari03}S. Onari, R. Arita, K. Kuroki, and H. Aoki,
cond-mat/0312314. 
\bibitem{comment1}We have not employed the projector QMC method because
the negative sign problem becomes too serious.
\bibitem{DMFA}A. Georges, G. Kotliar, W. Krauth, and M. J. Rozenberg,
Rev. Mod. Phys. {\bf 68}, 13 (1996).
\bibitem{Hirsch}J. E. Hirsch and R. M. Fye, Phys. Rev. Lett. {\bf 56}, 
2521 (1986).
\bibitem{comment2}As discussed by S. Moukouri, C. Huscroft, 
and M. Jarrell [Proceedings of the XIII Workshop on Recent Developments 
in Computer Simulation Studies in Condensed Matter Physics
(Springer 2000)], DCA gives a reliable
results even in purely one-dimensional systems if we take a 
sufficiently large cluster.
\bibitem{commentpxy} $p_{x+y} \sim \sin (k_x \pm k_y)$ 
may also be expressed as 
$p_{x+y} \sim \sin (k_x)\cos (k_y) \pm \cos (k_x)\sin (k_y)$, where 
$\sin (k_x)\cos (k_y)$ is the symmetry considered in \protect\cite{Nomura00}.
\bibitem{comment3}While it has been shown in Ref. \cite{Jarrell} 
that the sign problem is not serious for nearly half-filled cases, 
the problem becomes serious for lower densities, especially in 
the presence of $V$.
\bibitem{Bickers89}N.E. Bickers, D.J. Scalapino, and S.R. White,
Phys. Rev. Lett. {\bf 62}, 961 (1989).
\bibitem{Esirgen97}G. Esirgen and N.E. Bickers, Phys. Rev. B
{\bf 55}, 2122 (1997).
\bibitem{comment4}This is consistent with Ref.\cite{Kuroki01}, i.e., 
the pairing interaction for triplet pairs can be stronger than 
that for singlet pairs when $V$ is large enough to make charge 
fluctuations dominant.
\bibitem{commentd} The four-fold symmetry is slightly degraded 
in the $d_{xy}$ gap function due to a mixing with $d_{x^2-y^2}$.
\end{references}
\end{document}